\documentclass[draft]{agujournal2019}
\usepackage{url} 
\usepackage{lineno}
\usepackage[inline]{trackchanges} 
\usepackage{soul}
\usepackage{siunitx}
\DeclareSIUnit\year{year}
\DeclareSIUnit\years{years}
\DeclareSIUnit\ppm{ppm}
\draftfalse

\journalname{JGR: Machine Learning and Computation}
\begin{document}

\title{Stochastic Emulation of a Fully Coupled Preindustrial E3SMv3 Simulation}
\authors{Elynn Wu\affil{1}, James P. C. Duncan\affil{1}, Troy Arcomano\affil{1}, Jeremy McGibbon\affil{1}, Oliver Watt-Meyer\affil{1}, Christopher S. Bretherton\affil{1}, Naser Mahfouz\affil{2}, Claudia Tebaldi\affil{2}, Luke Van Roekel\affil{3}, Andrew Roberts\affil{3}, Wuyin Lin\affil{4}, Finn Rebassoo\affil{5}, Jean-Christophe Golaz\affil{5}, Peter M. Caldwell\affil{5}} 

\affiliation{1}{Allen Institute for Artificial Intelligence (Ai2), Seattle, WA, USA}
\affiliation{2}{Pacific Northwest National Laboratory, Richland, WA, USA}
\affiliation{3}{Los Alamos National Laboratory, Los Alamos, NM, USA,}
\affiliation{4}{Brookhaven National Laboratory, Upton, NY, USA}
\affiliation{5}{Lawrence Livermore National Laboratory, Livermore, CA, USA}

\correspondingauthor{Elynn Wu}{elynnw@allenai.org}
\begin{keypoints}
\item We develop a stochastic coupled atmosphere–ocean emulator of E3SMv3 that reproduces the model's mean climate state.
\item Stochastic training sustains realistic long-timescale variability, including ENSO, SST, and sea ice variability better than a deterministic baseline.
\item The emulator accurately captures precipitation statistics but underestimates the rarest tropical extreme events.
\end{keypoints}
\section*{Abstract}
We present a stochastic coupled emulator of E3SM version 3, built on the SamudrACE framework, which couples an atmosphere emulator (ACE2) with a full-depth ocean emulator (Samudra). We replace the deterministic atmosphere emulator with its stochastic counterpart, ACE2S, and fine-tune the coupled system with a probabilistic objective, so that the atmosphere acts as a source of internal variability for the ocean. Trained on 105 years of a pre-industrial control simulation and evaluated on an independent 400 years, the emulator reproduces E3SMv3's mean climate state with biases much smaller than existing model-to-observation differences. Relative to a deterministic baseline, stochastic training maintains internal variability across timescales, most notably in the ENSO power spectrum, eddy-rich SST anomalies, and sea ice variability in the marginal ice zone. The emulator captures daily precipitation accurately up to the 99.99th percentile, but underestimates the rarest tropical extremes. These results show that stochastic coupled emulators can reproduce long-timescale variability with high fidelity, while extrapolation to unseen extremes remains a key challenge.

\section*{Plain Language Summary}
Climate models are powerful tools for understanding how the Earth's climate works and how it may change, but they are extremely expensive to run, often requiring large supercomputers and long computing time. In recent years, researchers have started using artificial intelligence to build much faster emulators that learn to mimic these models at a tiny fraction of the cost. Most emulators so far have focused on a single part of the Earth system, such as the atmosphere or the ocean. In this study, we build an emulator that connects the atmosphere and the ocean together so they can influence each other, much like they do in the real world and in full climate models. A key feature of our approach is that it includes an element of randomness, which allows the emulator to reproduce the natural, unpredictable fluctuations of the climate. We find that our emulator closely reproduces the long-term behavior of a state-of-the-art climate model, including patterns like El Niño, while running roughly 40 times faster on a single graphics processor, whereas the original model requires thousands of processor cores. However, it struggles to reproduce the rarest and most extreme rainfall events. This kind of fast, accurate emulator could make it far easier to study climate variability and run the many simulations needed to understand our changing planet.

\section{Introduction}
General circulation models (GCMs) are our most advanced tools for understanding and projecting the Earth's changing climate. GCMs dynamically couple the atmosphere, ocean, sea ice, and land surface to capture the interactions of the Earth system components. Because of their physical and mathematical complexity, GCMs are computationally expensive. This limits the number of ensemble members that modeling centers generate for the Coupled Model Intercomparison Project (CMIP) \cite{dunne_evolving_2025, Eyring2016}. It also slows down GCM improvement; it often takes years for updates to be successfully integrated into a new model version.

Recent developments in AI models for weather prediction have proven highly successful, producing skillful short-term weather forecasts at drastically reduced computational cost \cite{Bi2023, Lam2023, Chen2024, Kochkov2024}. Data-driven global atmosphere models capable of multidecadal simulations \cite{watt2024ace2, Kochkov2024, chapman2025camulator} have also matured rapidly, as seen in the AI-based Atmosphere Model Intercomparison Project (AIMIP) \cite{henn_aimip_2026} --- an initiative that adapts a standard experimental protocol for traditional atmospheric models specifically for AI-driven atmosphere models. Data-driven models of other Earth system components have achieved comparable progress. In the ocean, these span from reanalysis-trained, eddy-resolving short-term forecasting systems that outperform operational numerical models \cite{xiong2023ai, wang2024xihe, aouni_glonet_2025} to climate-scale ocean emulators stable over century-long simulations \cite{dheeshjith_samudra_2025}. Data-driven sea ice models have progressed from regional Arctic surrogates with skill to seasonal timescales \cite{durand_data-driven_2024} to generative and mass-conserving global models that generalize across climates for decades \cite{finn_generative_2025, gregory_floenet_2026}. Land surface emulators using both statistical and deep learning approaches have demonstrated accurate, uncertainty-aware emulation of models such as JULES and ecLand at a fraction of the computational cost \cite{baker_emulation_2022, wesselkamp_advances_2025}.

To capture the full complexity of the Earth system, extending these data-driven techniques to fully coupled GCMs is the natural next step. Several attempts have been made, such as coupling an atmosphere emulator with a slab ocean model \cite{clark_ace2-som_2025}, combining data-driven atmosphere models with learned ocean components that predict only sea surface temperature \cite<SST; >{cresswell-clay_deep_2025} or the upper ocean \cite{wang2024xihe}, or coupling atmosphere and full-depth ocean emulators \cite{duncan_samudrace_2026}. The ultimate goal of such fully coupled emulators is to facilitate CMIP-style simulations in which systematic oceanic warming and heat storage must be reliably projected, and to generate the large ensembles that sample internal variability much more efficiently than the underlying GCM, supporting rapid model iteration and the exploration of climate scenarios that would be cost-prohibitive with conventional GCMs. 

Building on the recent success of SamudrACE \cite{duncan_samudrace_2026} --- which couples the ACE2 atmospheric emulator \cite{watt2024ace2} with the Samudra full-depth ocean emulator \cite{dheeshjith_samudra_2025} --- we develop a stochastic coupled emulator of E3SM version 3 \cite{golaz_energy_2026}, training on 105 years of a pre-industrial control simulation and evaluating on an independent 400-year segment of the same simulation. We call this emulator SamudrACE-E3SMv3.  The design and training of SamudrACE-E3SMv3 generally follow SamudrACE, but extend it in two ways. First, we replace the deterministic ACE2 atmosphere with its stochastic counterpart, ACE2S \cite{perkins_hiro-ace_2026}, and fine-tune the coupled system with a probabilistic objective. The stochastic atmosphere acts as a source of uncertainty for the ocean, so that SamudrACE-E3SMv3 becomes stochastic as well. Second, SamudrACE was originally trained on a 200-year pre-industrial simulation of GFDL's CM4 GCM.  Applying the SamudrACE framework to a different reference GCM serves as an important robustness test, helping to isolate whether emulator behaviors and limitations are intrinsic to the framework or artifacts of the particular GCM used for training. The 400-year evaluation period also substantially exceeds that available in the original SamudrACE work, enabling a more stringent assessment of long-term fidelity --- a prerequisite for the broader goal of reliable climate emulation. Skill is required not just in the mean state but also in variability and extremes across a range of timescales, a regime where stochastic emulators are expected to be particularly advantageous. We evaluate the resulting emulator against 400-year climatological means, precipitation distributions, and low-frequency variability. Emulation of E3SMv3's response to time-varying anthropogenic radiative forcings involves additional challenges and will be addressed in a follow-on paper.

\section{Data and Methods}
\subsection{E3SMv3 Data}
We start from a pre-existing 500-year pre-industrial control simulation of E3SMv3 \cite{golaz_energy_2026}, holding out the first 400 years as independent evaluation data. We re-run from 0401-01-01 to generate the additional output fields necessary to train SamudrACE--- 6-hourly outputs for the atmosphere and 5-day outputs for the ocean and sea ice. Of the 105-year training dataset, we train on the first 90 years and use the next 5 years for validation. The remaining 10 years are held out for testing. This reference E3SMv3 simulation is performed on an E3SM project cluster (AMD EPYC 7532, 64 processors per node), achieving about 28 simulated years per day using 105 nodes.

\subsection{SamudrACE Design}
SamudrACE consists of two component emulators --- atmosphere and ocean. For SamudrACE-E3SMv3, the atmosphere emulator is ACE2S \cite{perkins_hiro-ace_2026}, a stochastic variant of ACE2 \cite{watt2024ace2}, which operates with a 6-hourly time step, 1$^\circ$ horizontal resolution, and 8 vertical levels. ACE2S produces an ensemble of equally likely atmospheric states and is trained with a probabilistic loss combining the continuous ranked probability score (CRPS) and the energy score. The ocean emulator is Samudra \cite{dheeshjith_samudra_2025, yuan_samudra_2026}, which operates with a 5-day time step, 1$^\circ$ horizontal resolution identical to the atmosphere, and 19 vertical levels. The version used in SamudrACE also predicts sea ice concentration and volume in each grid column, although strict mass conservation is not imposed. To couple the emulators, the atmosphere provides 5-day mean surface fluxes, precipitation, and wind stress as boundary forcing to the ocean, and the ocean passes SST and sea ice fraction to the atmosphere. Table S2 lists all variables exchanged between ACE2S and Samudra. The ocean, sea-ice, and land fractions within each grid cell are forced to be consistent between the two emulators.

Training proceeds in two phases, broadly following \citeA{duncan_samudrace_2026}: the component emulators are first pretrained separately using perfect forcings --- boundary forcings taken directly from the E3SMv3 target data rather than from the other emulator --- then coupled and fine-tuned together. Unlike \citeA{duncan_samudrace_2026}, our coupled fine-tuning is stochastic and uses a single stage in which both components are optimized jointly. The following subsections describe the data pre-processing, uncoupled pretraining, and coupled fine-tuning.

\subsubsection{E3SMv3 Data Pre-processing}
Proper pre-processing of the E3SM output is a critical first step. For the atmosphere, we follow the procedures established by \citeA{duncan_application_2024} and \citeA{wu_applying_2025}. For the ocean and sea ice, the processing largely follows \citeA{duncan_samudrace_2026}. One important difference is that we perform most processing online while E3SM is running, rather than as an offline post-processing step. The raw atmosphere and ocean model output are conservatively coarsened in the vertical and remapped to a 1° Gaussian grid. For the ocean, to better align with E3SM's native vertical coordinate, we utilize a modified set of 19 vertical levels compared to \citeA{duncan_samudrace_2026}: [10, 25, 35, 45, 65, 95, 125, 155, 200, 320, 470, 775, 1050, 1400, 1850, 2400, 3100, 4000, 5500] m. Any grid cell with a non-zero ocean fraction is retained as a valid point by default. Notably, we omit the additional horizontal filtering employed in previous studies \cite{duncan_samudrace_2026, dheeshjith_samudra_2025}. Sea ice data is processed concurrently with the ocean data, as both share the same native grid.

We use the `aligned' setup described in \citeA{duncan_samudrace_2026} (Text S1) where we align ocean SST and sea ice state to the midpoint of the original 5-day averaging intervals rather than the endpoints. This ensures that surface forcings are applied consistently across both uncoupled pretraining and coupled fine-tuning. Saving snapshots of the E3SMv3 ocean state rather than 5-day means would circumvent the need for this temporal alignment but was not part of our original output specification.

\subsubsection{Atmosphere Pretraining}
ACE2S is pretrained with prescribed SST and sea ice following the protocol described in \citeA{perkins_hiro-ace_2026} and \citeA{clark_disentangling_2026}. Briefly, we split ACE2S training into two phases: the model is first trained stochastically with the CRPS and energy score loss over 1-step rollout for 30 epochs, then fine-tuned for a further 30 epochs over multi-step rollouts whose length is sampled per batch from \{1, 2, 4, 12, 20\} steps with probabilities \{0.6, 0.2, 0.1, 0.05, 0.05\}. In both stages, we use an ensemble size of two. During the second stage of stochastic fine-tuning, we additionally apply the total energy budget correction described in \citeA{clark_disentangling_2026}, which enforces conservation of total energy at each rollout step by imposing a vertically and horizontally uniform air temperature correction. In theory, the predicted global-mean total energy tendency should exactly balance the difference between the surface and top-of-atmosphere (TOA) energy fluxes; in practice, E3SMv3 exhibits a small residual imbalance of 0.09 W/m$^2$, which we impose as a constant unaccounted heating term in the correction. Unlike previous studies \cite{perkins_hiro-ace_2026, clark_disentangling_2026, duncan_samudrace_2026, wu_applying_2025, watt2024ace2}, we weight all variables equally in the loss, rather than using pre-determined per-variable weights.

\subsubsection{Ocean Pretraining}
Samudra is pretrained deterministically with prescribed atmospheric forcing, following \citeA{dheeshjith_samudra_2025}. We use the wider configuration described in \citeA{yuan_samudra_2026} which increases the channel widths in the encoder and decoder of the four ConvNeXt blocks to [280, 380, 480, 520]. This change helps reduce imprinting from deeper ocean depths onto the upper ocean. Pretraining proceeds in two stages: the model is first optimized over rollouts of four 5-day steps (20 days) for 150 epochs, then fine-tuned over rollouts of eight steps (40 days) for a further 20 epochs.

\subsubsection{Coupled Fine-tuning}
\begin{figure}
\noindent\includegraphics[width=\textwidth]{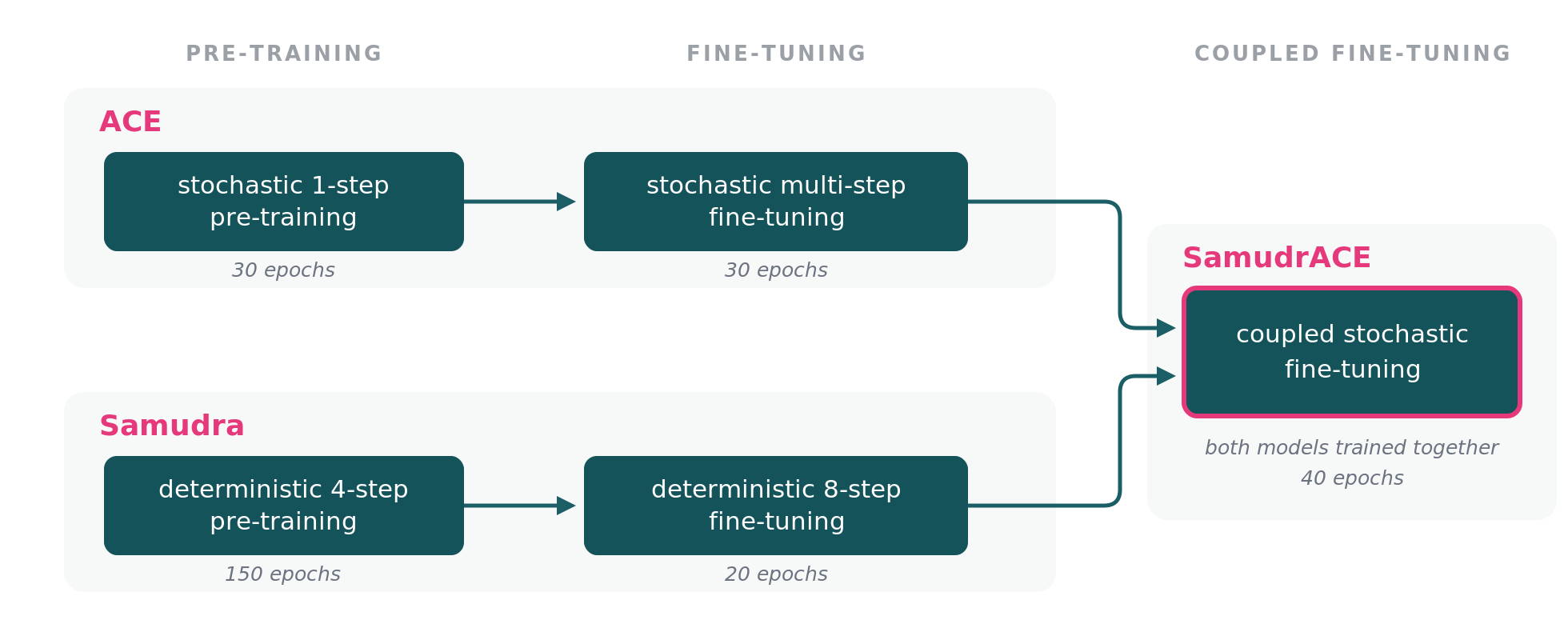}
\caption{Overview of the SamudrACE-E3SMv3 training protocol.}
\label{fig:training-flowchart}
\end{figure}

After pretraining, we couple ACE2S and Samudra and fine-tune both components jointly in a single stage for 40 epochs, in contrast to the two-stage approach of \citeA{duncan_samudrace_2026} in which atmosphere weights were initially frozen. Figure \ref{fig:training-flowchart} shows the overall training workflow. Each training sample is rolled out over four coupled steps (20 days, corresponding to four 5-day ocean steps and eighty 6-hourly atmosphere steps). 

A key consequence of coupling to a stochastic atmosphere is that Samudra can be trained stochastically as well. During fine-tuning, two ensemble members are generated for each training sample: ACE2S produces two realizations whose 5-day mean surface forcings drive two corresponding Samudra rollouts. Both components are optimized with the same probabilistic objective used for ACE2S pretraining, a weighted combination of CRPS and the energy score, with equal loss weights on the atmosphere and ocean components. ACE2S thereby acts as the source of uncertainty for the coupled ocean; no additional noise is injected into the Samudra model architecture. Further details on how the losses are computed and how coupled fine-tuning affects Samudra are described in Supplementary Material S1-S3.

To keep memory and computational cost manageable while still exposing the models to long rollouts, the number of forward steps included in each component's loss is sampled independently per batch: the ocean loss window is drawn from \{0, 1, 2, 4\} ocean steps with probabilities \{0.1, 0.3, 0.3, 0.3\}, and the atmosphere loss window from \{0, 1, 2, 4, 21, 41\} atmosphere steps (up to just over 10 days) with probabilities \{0.025, 0.29, 0.29, 0.29, 0.1, 0.05\}, where a window of zero steps means that component contributes no loss for that batch. The ocean is optimized at every step within its sampled window, while the atmosphere is optimized only at the final step of its window. The longer atmosphere windows (21 and 41 steps) extend beyond one and two ocean coupling steps, respectively, so that the atmosphere is evaluated after responding to one or two updated ocean states rather than a static one, exposing the loss to the coupled interaction.

To select the best model checkpoint, we conduct inline inference using eight 12-year segments initialized at regular intervals to span the full training dataset. The final checkpoint is chosen based on the lowest aggregate RMSE across all atmospheric and oceanic variables during inline inference.

For comparison, we also train a deterministic coupled emulator similar to the original SamudrACE. Its atmosphere component is the deterministic version of ACE2, and both components are fine-tuned jointly with an MSE loss using a single ensemble member. In place of the randomly sampled loss windows described above, the deterministic baseline uses fixed windows of four ocean steps (20 days) and two atmosphere steps (12 hours), matching the loss structure of the second fine-tuning stage in \citeA{duncan_samudrace_2026}. All other training settings, including the optimizer, learning rate schedule, and number of epochs, are identical to the stochastic configuration.

\subsection{SamudrACE-E3SMv3 Data}
We initialize SamudrACE-E3SMv3 at the beginning of the training period and run it for 500 years. The first 105 years (model years 0401--0505) overlap the training period. These in-sample years are compared against the corresponding E3SMv3 output. This is a nontrivial test since the training target was a 6-hour forecast and the evaluations are from a long simulation. For a fully independent evaluation, the subsequent 400 years (0505--0905) of SamudrACE-E3SMv3 output are compared against the E3SMv3 model years 0001--0400 that sample the same stationary pre-industrial climate but were not seen during training. For this period, only a subset of variables from the original E3SMv3 simulation was saved, at either daily or monthly frequency. On a single H100 GPU, SamudrACE-E3SMv3 achieves inference speeds of 1105 SYPD (stochastic) and 1360 SYPD (deterministic).

\section{Results}
\subsection{Climatology}
\begin{figure}
\centering\includegraphics[width=\textwidth]{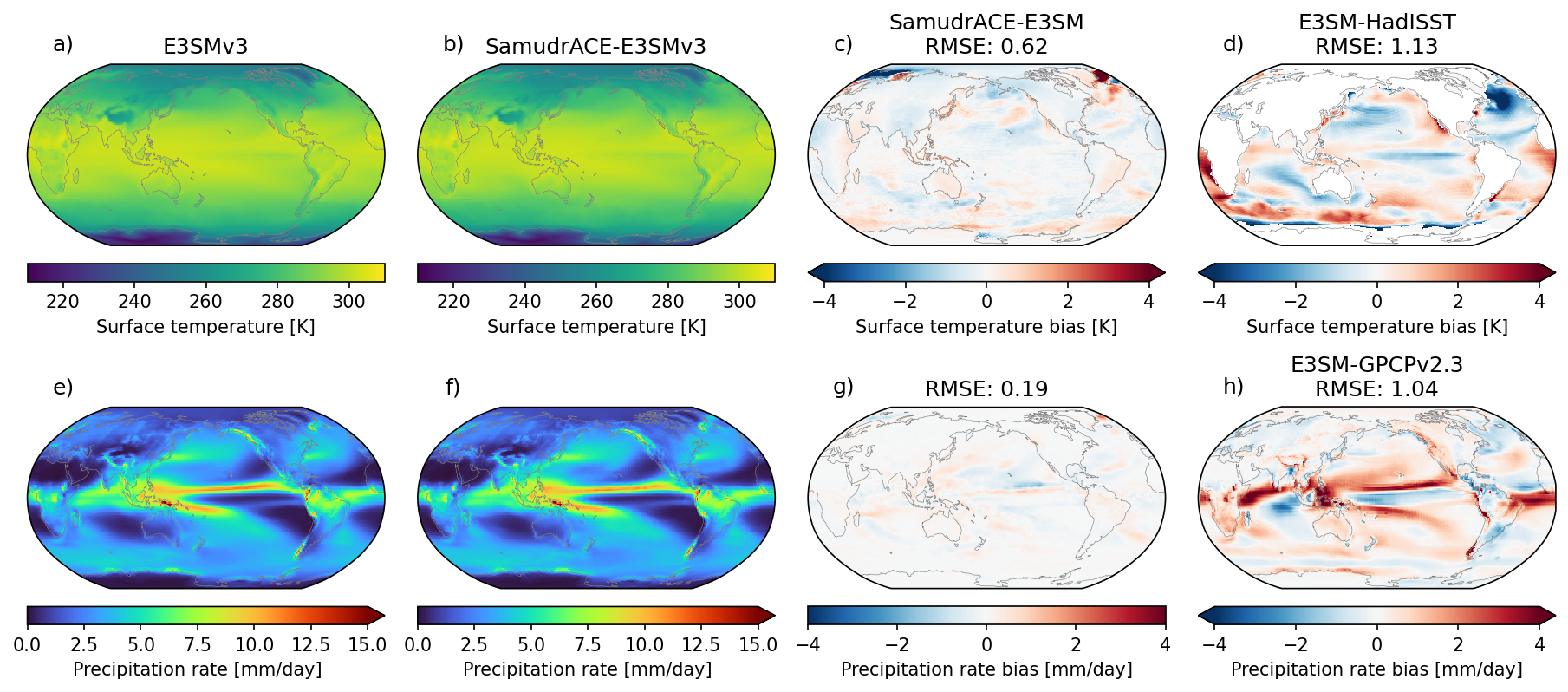}
\caption{400-year time mean of surface temperature and precipitation for E3SMv3 (a and e), SamudrACE-E3SMv3 (b and f), bias between SamudrACE-E3SMv3 and E3SMv3 (c and g), and bias between E3SMv3 and observation (d and h). For comparison with HadISST, we define ocean grid cells as ocean fraction greater than 96\unit{\percent} and only compare SST over these points. Note the observation datasets use different time spans that are specified in the text.}
\label{fig:climatology-maps}
\end{figure}

Consistent with the findings of \citeA{duncan_samudrace_2026}, SamudrACE-E3SMv3 accurately reproduces the spatial structure of the mean climate state of E3SMv3. We compare the time-mean biases between SamudrACE-E3SMv3 and E3SMv3 against the existing biases between E3SMv3 and available observation data (Figure \ref{fig:climatology-maps}). For surface temperature observation data, we use the Hadley Centre Sea Ice and Sea Surface Temperature dataset (HadISST) \cite{hurrell_new_2008} averaged over 1870--1900, the closest available period to pre-industrial conditions. For precipitation observations, we use the Global Precipitation Climatology Project v2.3 \cite{adler_global_2018} averaged over 1985–2014. SamudrACE-E3SMv3's root-mean square biases (RMSBs) vs. E3SMv3 (0.62~K, 0.19~mm~day$^{-1}$) are substantially smaller than  E3SM's against the observations (1.13~K, 1.04~mm~day$^{-1}$); i.e the emulator's added bias is small compared with E3SMv3's own bias from observations. 

A persistent time-mean surface temperature bias between SamudrACE-E3SMv3 and E3SMv3 is found in the marginal ice zone (MIZ) of the Barents and Greenland Seas (Figure S1). Surface temperature and turbulent fluxes respond nonlinearly to the sea ice partition here, so small biases in the mean ice edge — which sweeps across this sector between March and September (Figure S1a) — translate into large local surface temperature biases. This region is intrinsically difficult for the ocean emulator. The uncoupled Samudra run, forced by prescribed E3SMv3 fields, already generates disproportionate large SST biases in the Nordic Seas (Figure S1b), with the error maximizing at intermediate ice concentration and vanishing under consolidated pack ice. With prescribed (perfect) atmospheric forcing, an ice-edge error is continuously corrected. However, in the coupled system it alters the emulated surface fluxes and further displace the ice edge. The Nordic-box SST RMSB grows from 0.15 K uncoupled to 2.1 K coupled, and the coupled error organizes into a dipole: warm along the East Greenland outflow, cold in the northern Barents Sea (Figure S1c). This is reproducible across rollouts and random seeds used to initialize training.

Inconsistencies in the coupled training target compound this error. We take sea ice fraction from the E3SMv3 atmosphere output and SST, ocean state, and sea ice volume from the E3SMv3 ocean output. This was deliberate: we wanted the atmosphere emulator to train on the same sea ice fraction the physical atmosphere model saw, kept consistent between uncoupled and coupled training. In hindsight, using the ocean model's sea ice fraction would have been more consistent, though the effect on the final result is unclear. Both products are 5-day averages, so the inconsistency is not temporal but one of source and grid. The atmosphere's sea ice fraction (on its native grid) and the ocean model's fields (on the ocean grid) disagree most near the ice edge, and since the atmosphere's training surface temperature blends ocean SST with the atmosphere's ice fraction, that disagreement enters the target precisely there.

\begin{figure}
\centering\includegraphics[width=\textwidth]{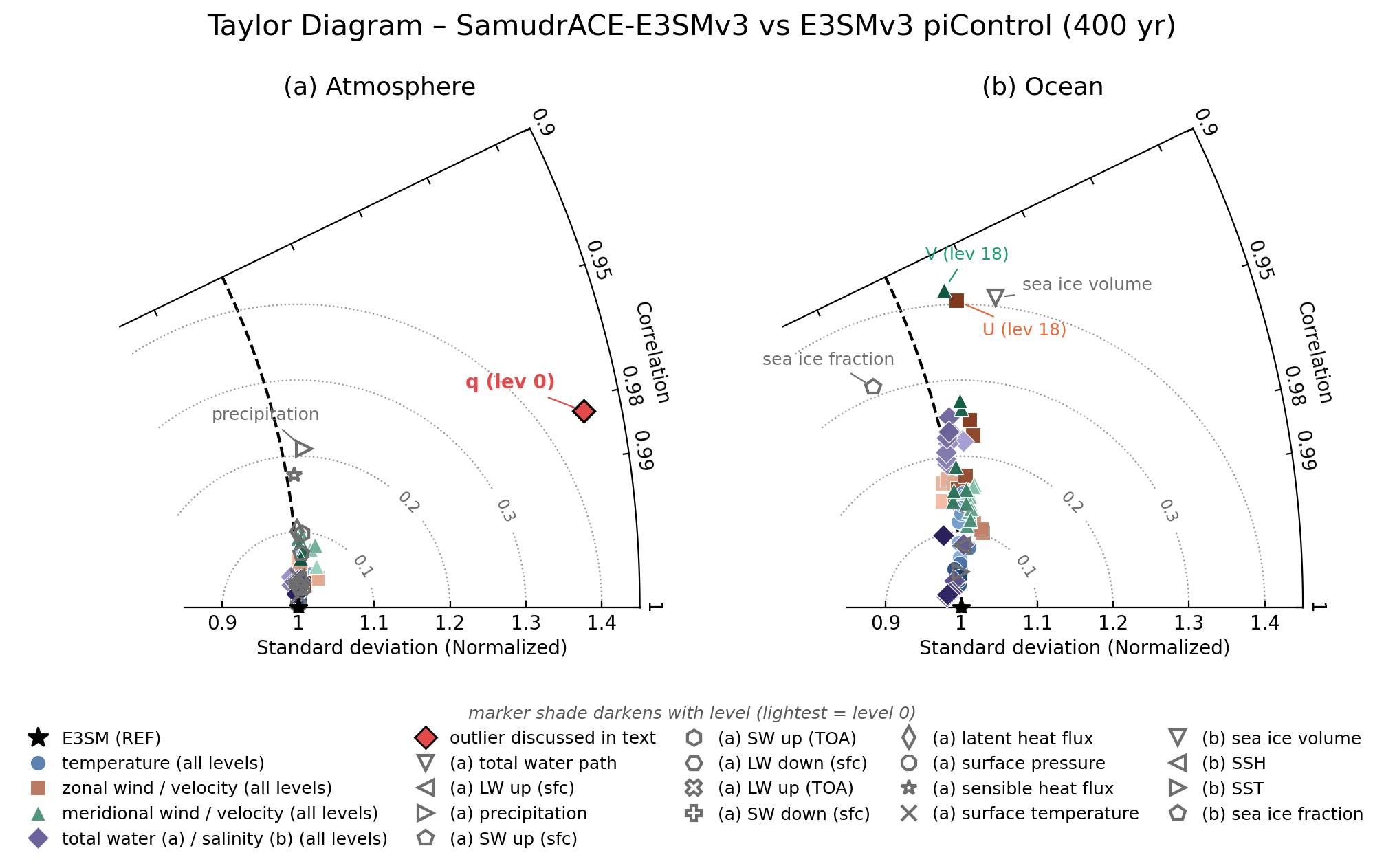}
\caption{Taylor diagram of 400-year time mean atmosphere (a) and ocean (b) variables between SamudrACE-E3SMv3 and E3SMv3. Colors distinguish variables; shading darkens with increasing level index, from levels 0–7 in the atmosphere and 0–18 in the ocean. Level 0 is the lightest shade, corresponding to the top of the atmosphere in (a) and the near-surface layer in (b). Total water in (a) is specific total water, which is the sum of water vapor, cloud liquid, cloud ice, and rain mixing ratio (kg/kg); falling snow is not included.}
\label{fig:taylor-diag}
\end{figure}

Figure \ref{fig:taylor-diag} shows a Taylor diagram \cite{yarger_autocalibration_2024, taylor_summarizing_2001} of all variables in the atmosphere and ocean over the 400-year evaluation period. Atmospheric vertical levels are indexed from the top of the atmosphere (level 0) downward, while ocean levels are indexed from the surface (level 0) downward. Here the standard deviation of the emulator field normalized by the standard deviation of the E3SMv3 field is the radius, and the correlation between the two fields is the angular coordinate. Almost all variables are concentrated near the zero-bias point at (1,0), with the exception of stratospheric specific total water (\texttt{q0}), whose near-uniform
time-mean field causes the normalized metrics to amplify a physically negligible error.

\subsection{Precipitation histogram}
\begin{figure}
\centering\includegraphics[width=1\textwidth]{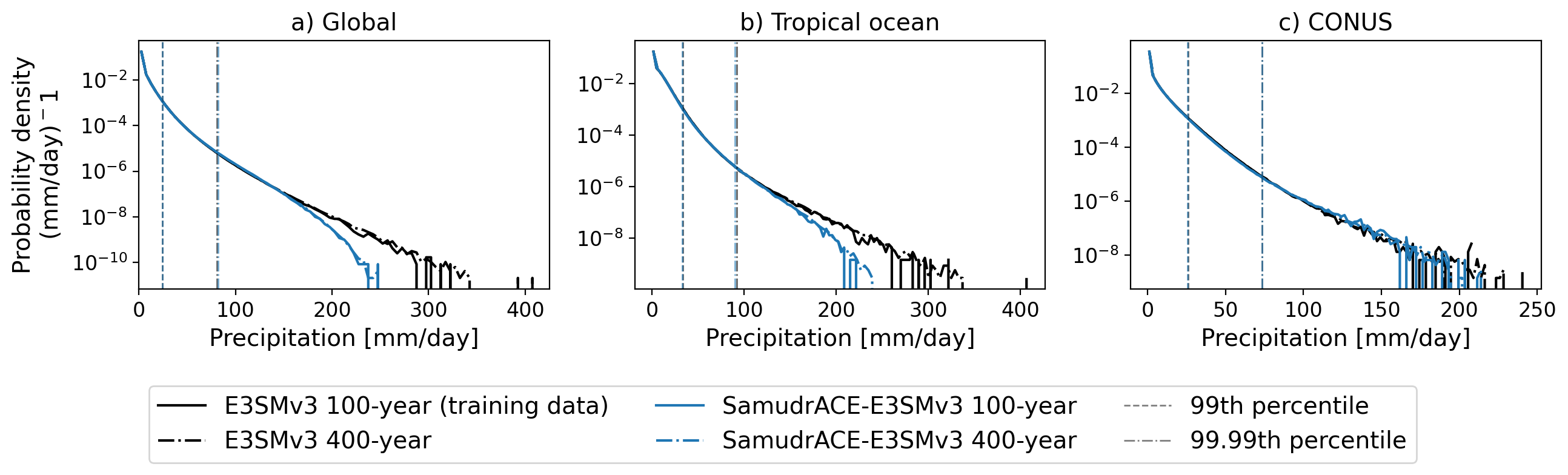}
\caption{Daily precipitation probability density over the (a) global, (b) tropical ocean, and (c) CONUS domains, comparing E3SMv3 with SamudrACE-E3SMv3. Solid E3SMv3 lines show the 100-year training-period data; solid SamudrACE-E3SMv3 lines show the first 100 years of a free-running rollout. Dash-dot lines show the 400-year period: for E3SMv3 this is independent evaluation data, and for SamudrACE-E3SMv3 it is the free-running rollout following the initial 100-year segment. Vertical lines mark the 99th and 99.99th percentiles of the evaluation period. Note that axis scaling are different for the three panels.}
\label{fig:precip_hist}
\end{figure}

Figure~\ref{fig:precip_hist} compares daily precipitation distributions of SamudrACE-E3SMv3 and E3SMv3 globally and for two contrasting regimes: the tropical ocean ($10^\circ$S--$10^\circ$N), dominated by intense, intermittent convection, and the Continental United States (CONUS), where precipitation is predominantly frontal and stratiform. For each model we overlay the 100-year training period and the 400-year evaluation period. While the full training segment spans 105 years, for simplicity we use its first 100 years for the in-sample comparison.

For E3SMv3, the two periods are nearly indistinguishable except in the far tail, where the longer record simply samples rarer events--- as expected for a stationary pre-industrial control climate. SamudrACE-E3SMv3 behaves the same way: its first 100 years and the subsequent 400-year free-running rollout produce essentially identical distributions without evidence of spurious climate drift.

SamudrACE-E3SMv3 reproduces the bulk of the daily precipitation distribution with high fidelity.  The 99th and 99.99th percentiles agree with E3SMv3 to within a few percent, even in the 400-year free-running rollout. Over the continental U.S. (CONUS), SamudrACE-E3SMv3 tracks E3SMv3 into the far tail in both periods, including the rarest events sampled in 400 years.

The discrepancy in the extreme tail of the global daily precipitation distribution comes from tropical convective events. Over the tropical ocean, SamudrACE-E3SMv3 underestimates the frequency of events above 150~mm~day$^{-1}$ and terminates near 230~mm~day$^{-1}$, whereas the E3SMv3 distribution extends past 300~mm~day$^{-1}$. Accurately emulating the frequency of such extreme precipitation events may require training strategies that explicitly up-weight the far tail, such as tail-weighted loss functions or up-sampling \cite{sun_can_2025}.

\subsection{El Niño-Southern Oscillation (ENSO)}
\begin{figure}
\noindent\includegraphics[width=\textwidth]{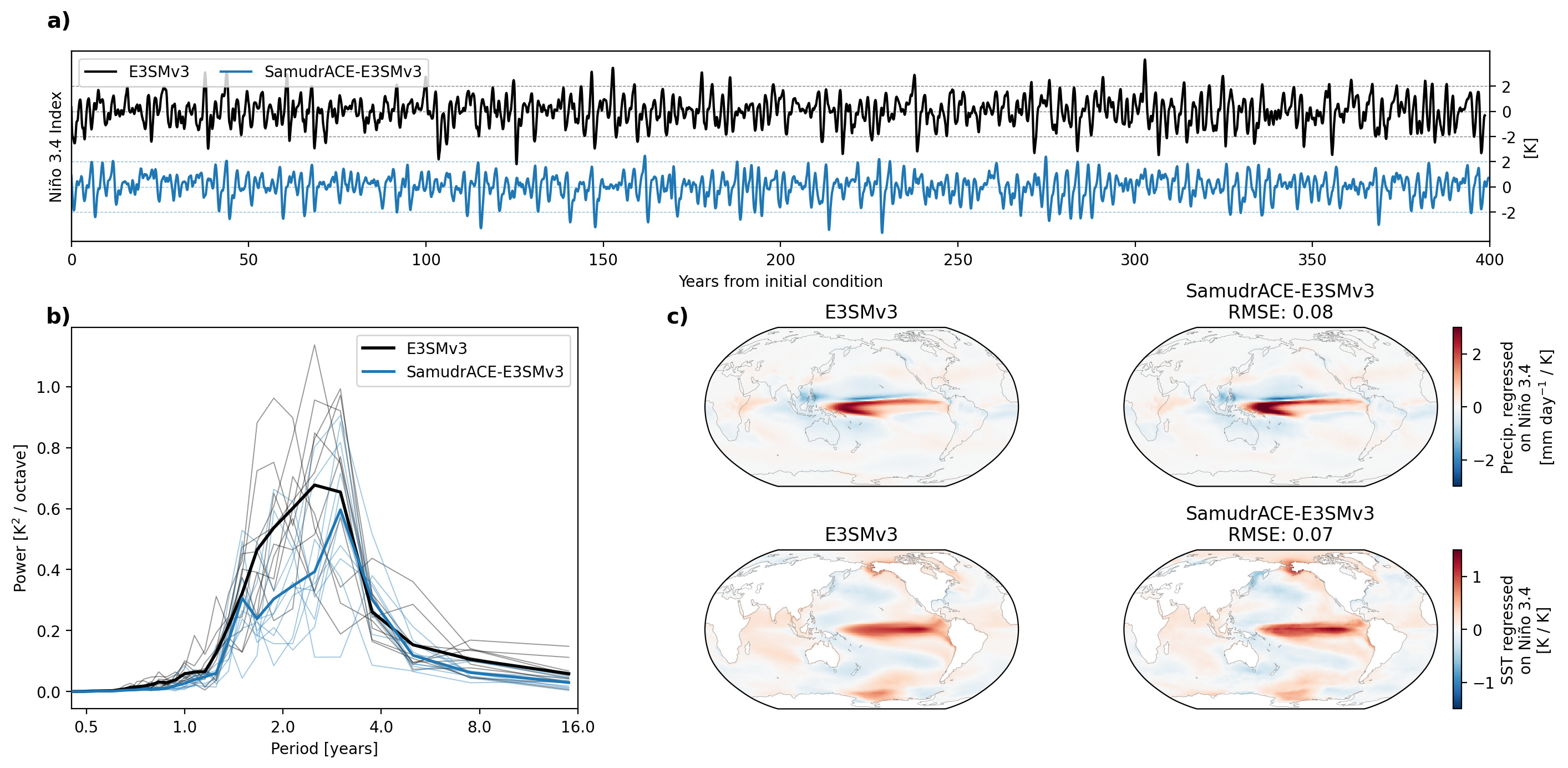}
\caption{ENSO characteristics in the 400-year E3SMv3 simulation and SamudrACE-E3SMv3 for (a) time series of monthly mean Niño 3.4 index; (b) corresponding power spectra, where thin lines show the spectra of individual 40-year blocks and thick lines their mean; (c) regression of precipitation (top) and SST (bottom) spatial patterns onto the Niño 3.4 index, for E3SMv3 and SamudrACE-E3SMv3. RMSE is quoted for SamudrACE-E3SMv3 relative to E3SMv3.}
\label{fig:enso}
\end{figure}

We evaluate the ENSO characteristics of SamudrACE-E3SMv3 against E3SMv3 in Figure \ref{fig:enso}. The Niño 3.4 index is computed identically for both models: surface temperature is area-averaged over the Niño 3.4 region (5$^\circ$S--5$^\circ$N, 170$^\circ$W--120$^\circ$W), anomalies are taken with respect to the full monthly climatology, and a 5-month running mean is applied. The resulting 400-year index time series (Figure \ref{fig:enso}a) shows that the emulator maintains stable, realistic interannual variability comparable to E3SMv3, with no drift or collapse in variance over the entire 400-year rollout.

Figure \ref{fig:enso}b compares the Niño 3.4 power spectra. To characterize sampling uncertainty, each 400-year index is split into ten independent, non-overlapping 40-year blocks, with the overall mean removed. Within each block, the power spectral density is estimated using Welch's method with 15-year Hann-windowed segments at 50\% overlap, yielding one smoothed spectrum per block (shown as thin lines), and the bold lines show the average across all blocks. The block-to-block spread within E3SMv3 itself is substantial: spectral power at the ENSO peak varies by roughly a factor of four across 40-year blocks of the same control simulation. This spread reflects the combined effect of sampling uncertainty in 40-year spectral estimates and any genuine low-frequency modulation of ENSO amplitude, and defines the envelope against which the emulator should be judged: a single short record is insufficient to characterize ENSO behavior in either model. Against this benchmark, SamudrACE-E3SMv3 reproduces the dominant ENSO peak and sustains power in the low-frequency tail at periods beyond 4 years, with its block spectra largely overlapping E3SMv3's block spread, though with a slight low bias. However, SamudrACE-E3SMv3 underestimates E3SMv3's spectral power by up to 50\% at shorter interannual periods (roughly 1.5–2 years) and concentrates its variance in a somewhat narrower peak near 3 years, whereas E3SMv3's peak is broader and extends to shorter periods. The ratio of block-to-block spread to time-mean spectral power in SamudrACE-E3SMv3 is comparable to E3SMv3 in all spectral bands, suggesting the coupled emulator captures the intrinsic randomness of ENSO simulated by E3SMv3. 

The emulated Niño 3.4 power spectrum is somewhat more faithful to the reference model than found in our previous paper \citeA{duncan_samudrace_2026}, which used a deterministic SamudrACE emulator of GFDL-CM4. That study reported a Niño 3.4 spectral peak stronger and narrower than the target's, with power deficits at both the 2-year shoulder (as seen here) and periods beyond 4 years (larger than seen here).  

We attribute the more realistic ENSO variability in SamudrACE-E3SMv3 to the use of a stochastic rather than deterministic atmospheric emulator. To isolate this effect, we compare emulators trained with two deterministic and two stochastic random seeds. For each seed selected, we show the checkpoint with the best long-rollout climate skill, our standard model-selection criterion, which is independent of any ENSO diagnostic. Figure~S2 compares the resulting 400-year Niño 3.4 indices and power spectra. One deterministic seed produces a plausible spectrum, but the other collapses onto an overly regular oscillation with spectral biases qualitatively similar, though more extreme, than reported by \citeA{duncan_samudrace_2026}. Its Niño 3.4 index is visibly periodic, its spectral peak near 3 years is significantly stronger than E3SMv3's, a secondary peak appears at the 1.5-year period, and power at periods beyond 4 years is largely absent. This behavior is shared by all checkpoints of that training seed (not shown), indicating a property of the training run rather than of the checkpoint. Both stochastic seeds instead produce irregular, correctly banded variability, with spectral peaks within a factor of about two of E3SMv3 across all checkpoints examined. Importantly, the collapsed deterministic model is indistinguishable from the others in time-mean climate skill--- its 400-year surface temperature and precipitation pattern RMSEs (0.66~K and 0.37~mm~day$^{-1}$) are similar to those of the plausible deterministic seed (0.67~K and 0.41~mm~day$^{-1}$), suggesting that mean-state metrics alone cannot detect this failure mode. 

Finally, Figure \ref{fig:enso}c shows the regression of global precipitation and sea surface temperature anomalies onto the Niño 3.4 index. SamudrACE-E3SMv3 closely matches E3SMv3's precipitation teleconnection pattern, including the eastward shift and intensification of precipitation over the central Pacific warm pool and the compensating drying over the Maritime Continent and subtropics. The accompanying SST regression reproduces the canonical ENSO structure: a warm anomaly tongue spanning the central and eastern equatorial Pacific and a weaker cool anomaly over the western and off-equatorial Pacific. The subsurface response is consistent with this picture. Regression of ocean potential temperature onto the Niño 3.4 index along the equatorial Pacific (Figure S3) shows that SamudrACE-E3SMv3 reproduces E3SMv3's thermocline dipole, with warm anomalies in the eastern Pacific and compensating cool anomalies centered near 150-200 m in the west. The principal discrepancies are a surface warm bias in the far eastern Pacific and a weaker positive bias near 100 m in the central Pacific. Together these results suggest that the coupled emulator has learned the correct ENSO response: the ocean component generates the proper large-scale SST anomaly pattern associated with a given Niño 3.4 state, and the atmospheric component simulates the corresponding precipitation teleconnection with high spatial fidelity relative to E3SMv3.

\subsection{Internal variability: stochastic vs. deterministic}
\begin{figure}
\noindent\includegraphics[width=\textwidth]{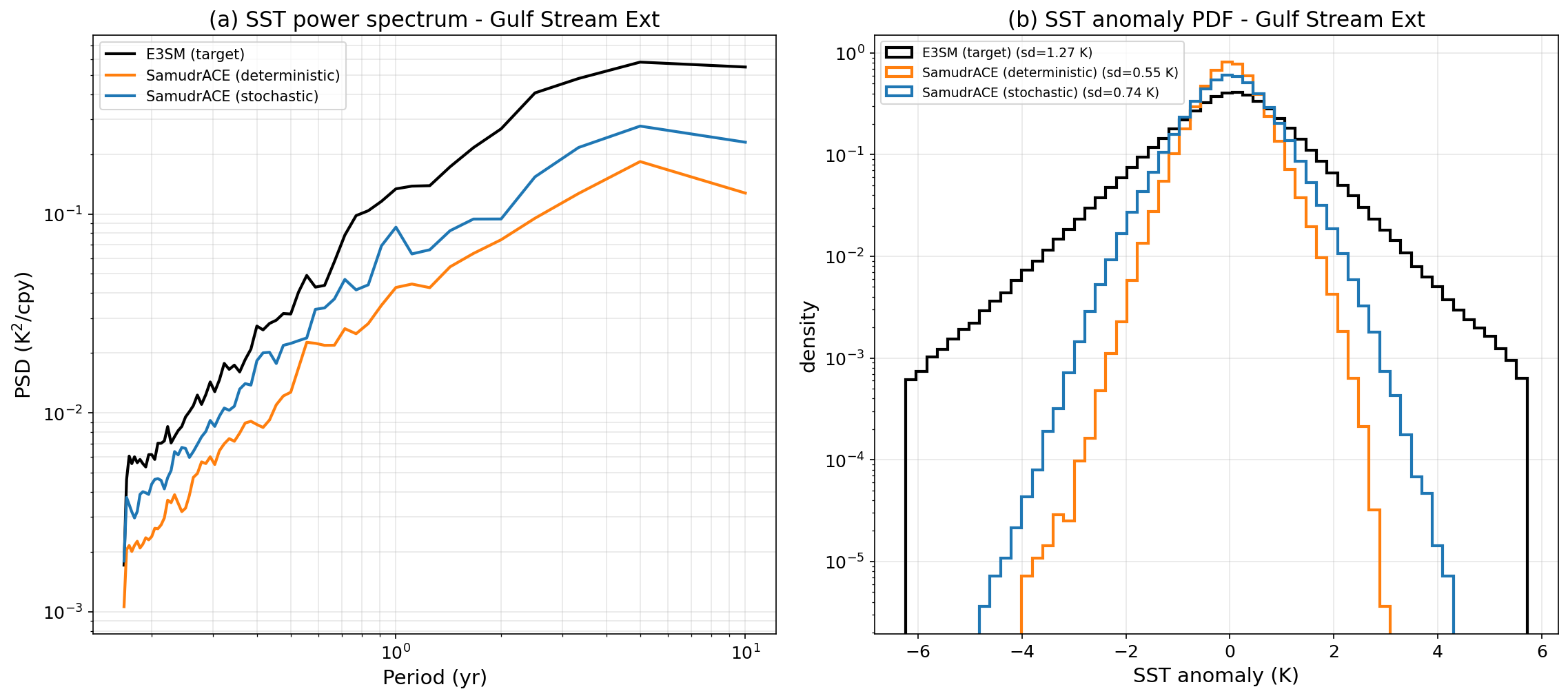}
\caption{Sea surface temperature variability in the Gulf Stream Extension region ($35^\circ$--$45^\circ$N, $75^\circ$--$45^\circ$W) over the 400-year evaluation. Comparison of the E3SMv3 target, the deterministic SamudrACE emulator, and the stochastic SamudrACE emulator for (a) power spectral density of the box-mean SST anomaly index, and (b) probability density of pooled SST anomalies. Anomalies are deseasonalized by removing the climatological monthly cycle, and power spectra are computed with Welch's method (10-yr segments) and averaged over samples.}
\label{fig:det-vs-stochastic}
\end{figure}

SST variability in the Gulf Stream Extension provides evidence that a stochastic emulator is needed to fully represent internal ocean variability. The deterministic emulator systematically damps SST fluctuations in this region relative to the E3SMv3 target. At the gridpoint scale, the standard deviation of deseasonalized monthly SST anomalies falls from 1.27 K in the target to 0.55 K in the deterministic run, and to 0.74 K in the stochastic run (Figure \ref{fig:det-vs-stochastic}b). The damping is not a grid-scale effect as the box-mean SST index retains 60\% of the per-gridpoint anomaly amplitude in the target, indicating variability that is coherent across the box, and the same deficit appears in the index itself (0.74 K target, 0.38 K deterministic, 0.49 K stochastic). The power spectrum of the box-mean index is a factor of 2.5--4 below E3SMv3 across the resolved band from sub-annual to decadal periods, while the stochastic run closes roughly half that gap (Figure \ref{fig:det-vs-stochastic}a). This is consistent with a deterministic, MSE-minimized model regressing toward its conditional mean and suppressing the fluctuations it cannot predict. 

Adding stochasticity to the emulator recovers much of this missing variance, especially at shorter timescales at which there is stronger weather-related variability in ocean surface wind stress and heat flux forcing to help drive mesoscale ocean eddies. The anomaly standard deviation rises 34\% from 0.55 K to 0.74 K, and the power spectrum tracks the E3SMv3's shape and amplitude more closely across nearly the entire resolved band (Figure \ref{fig:det-vs-stochastic}a,b) though a gap to E3SMv3 remains. A similar pattern, in which the stochastic model captures more of E3SMv3's mesoscale ocean variability, also holds for SSH, SST, and surface kinetic energy in this eddy-rich region (Figures S4 and S5).

\begin{figure}
\noindent\centering\includegraphics[width=0.9\textwidth]{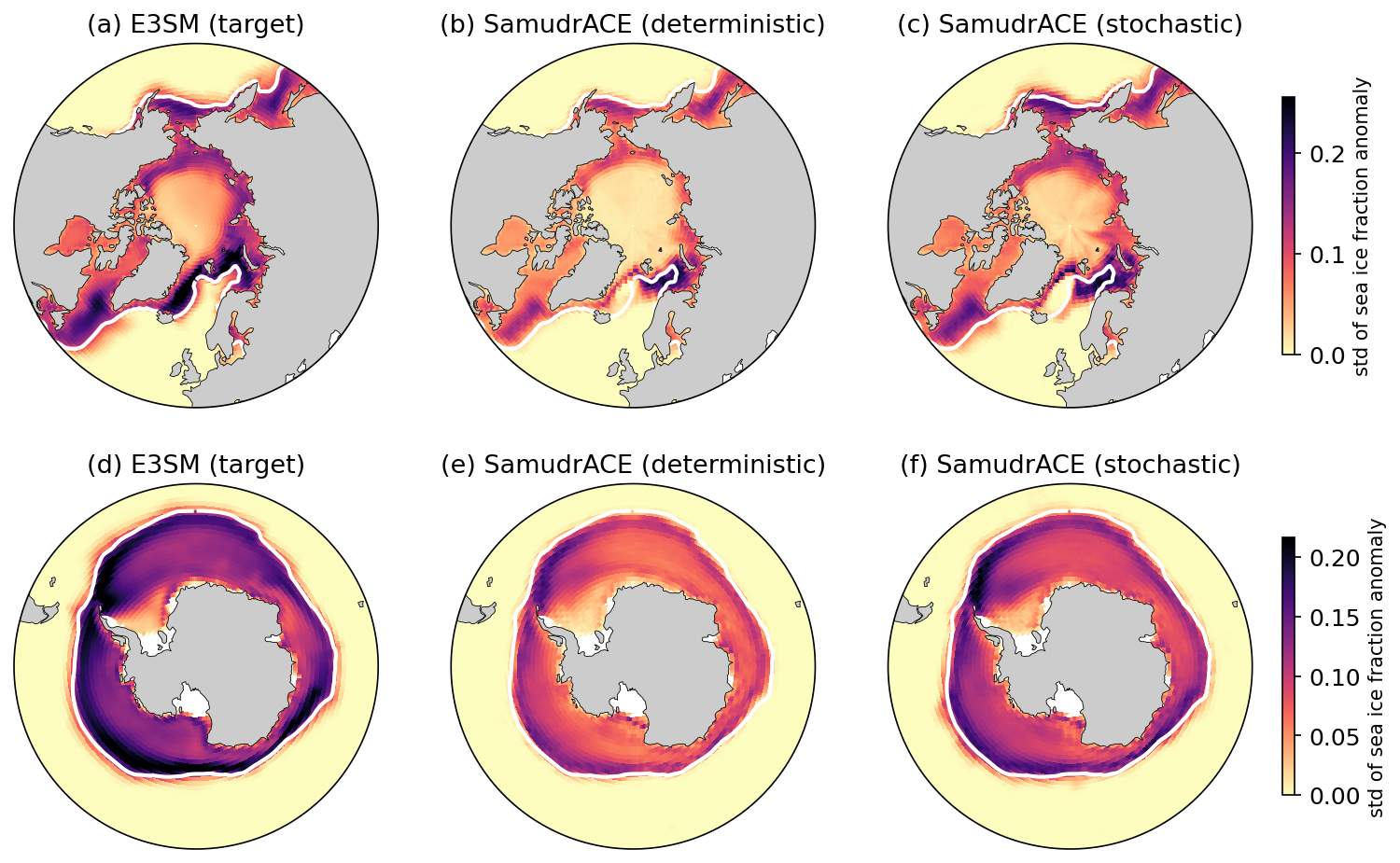}
\caption{Local sea-ice variability in the Northern (a-c) and Southern (d-f) Hemisphere: standard deviation of deseasonalized monthly sea-ice fraction anomalies over the 400-year evaluation period for the E3SMv3 target (a, d), the deterministic SamudrACE emulator (b, e), and the stochastic SamudrACE emulator (c, f). The white contour marks E3SMv3's climatological-maximum 15\% concentration.}
\label{fig:det-vs-stochastic-sic}
\end{figure}

Maps of local sea-ice fraction variability (Figure \ref{fig:det-vs-stochastic-sic}) show that both emulators underestimate E3SMv3's variability throughout the marginal ice zone (MIZ), and that stochastic training reduces but does not eliminate this shortcoming. In both hemispheres, E3SMv3's variability forms a ring along the seasonally migrating ice edge, peaking near the climatological-maximum 15\% concentration contour, where the ice margin advances and retreats from month to month. Both emulators reproduce this geographic pattern but with systematically weaker amplitude everywhere. The deterministic emulator reduces the MIZ-averaged anomaly standard deviation to 0.52$\times$ the target in the Northern Hemisphere and 0.57$\times$ in the Southern Hemisphere. The stochastic emulator improves these ratios to 0.70$\times$ and 0.76$\times$.

\section{Conclusions}
We present SamudrACE-E3SMv3, a fully coupled emulator trained on 105 years of the E3SMv3 pre-industrial control simulation and evaluated on 400 years of independent data. The model demonstrates high fidelity in reproducing E3SMv3's mean climate state. The biases between SamudrACE-E3SMv3 and E3SMv3 are significantly smaller than the differences between E3SMv3 and present-day observational data for SST and precipitation. A regional exception to this is a persistent emulator bias near the Barents Sea, concentrated in the marginal ice zone where surface fluxes and temperature respond nonlinearly to small errors in the sea ice partition. This bias is likely compounded by spatial and temporal inconsistencies among the atmosphere, ocean, and sea ice training data, and by SST biases inherent to the uncoupled Samudra ocean emulator in this region.

SamudrACE-E3SMv3 reproduces the daily precipitation distribution with high fidelity across the global, tropical ocean, and CONUS domains, matching E3SMv3 out to the 99.99th percentile under a 400-year free-running rollout. The emulated precipitation climate shows no drift: the first 100 years of the rollout and the subsequent 400 years produce essentially identical distributions, mirroring the stationarity of the E3SMv3 control climate itself. Over CONUS, the agreement extends into the far tail, including the rarest events sampled in 400 years. The one shortfall is the extreme tail of tropical convection, where the emulator underestimates the frequency of the most intense daily events; this low bias is equally present in the training and evaluation periods, and addressing it may require bespoke training strategies.

Stochastic SamudrACE-E3SMv3 maintains low-frequency variability on multi-year timescales better than the deterministic baseline. This is evident in its representation of ENSO, where stochastic SamudrACE-E3SMv3 captures the power spectrum of the Niño 3.4 index well, especially for periods longer than 4 years, while deterministic training is prone to overly regular, spectrally concentrated variaiblity, in one case losing most power at periods beyond 4 years. A similar improvement appears in the variability of the coupled ocean and sea ice. In the Gulf Stream Extension, stochastic SamudrACE-E3SMv3 recovers some of the SST variance damped by the deterministic emulator, though it still falls short of the target's amplitude. The same holds for sea ice variance. 
The stochasticity originates in noise injected through the ACE2S architecture, and during coupled fine-tuning both components are optimized with the same probabilistic objective. Taken together, these results indicate that stochasticity enhances the coupled system's internal variability and moves it closer to the reference, with the deterministic ocean inheriting its variability from the stochastic atmosphere.

While there are still many challenges to increasing the capabilities of fast AI emulators of full earth system models, our rapid progress to date is encouraging. Future research will extend SamudrACE-E3SMv3 to generalize over historical simulations with changing greenhouse gases and aerosols, investigating the mechanisms by which stochastic training improves the recovery of internal variability, and evaluating whether a single stochastic emulator can reproduce the ensemble statistics of a large ensemble of fully coupled historical simulations. It would also be physically reasonable to add noise internal to Samudra to represent stochastic feedbacks from unresolved ocean scales onto the resolved-scale currents; the importance of such ocean-internal noise relative to atmosphere-forced stochasticity remains be explored. We are developing a mass and energy conserving sea ice emulator that could further bolster the physical interpretability, accuracy and generalizability of SamudrACE.  

\section*{Open Research Section}
Training data used in this manuscript can be downloaded via Guest Collections on Globus under SamudrACE-E3SMv3 at \url{https://app.globus.org/file-manager/collections/b79ba000-e615-4675-8121-725b6bbffaef/overview}. This data is hosted through NERSC SHARE. The code used for model training and evaluation is archived through Zenodo \cite{mcgibbon_ai2cmace_2026}, and the scripts used for generating figures are also archived through Zenodo \cite{wu_2026_21879164}. Additionally, the model checkpoint, initial conditions and forcing data can be downloaded from Hugging Face at \url{https://huggingface.co/allenai/SamudrACE-E3SMv3}.

\acknowledgments
Ai2 is supported by the estate of Paul G. Allen. This research was also supported as part of the Energy Exascale Earth System Model (E3SM) project, funded by the U.S. Department of Energy, Office of Science, Office of Biological and Environmental Research. The reference E3SM simulations and portions of SamudrACE training were completed using resources of the National Energy Research Scientific Computing Center (NERSC), a Department of Energy Office of Science User Facility using NERSC award. The remaining portions of SamudrACE training and evaluation were completed with Ai2 computational resources. We thank other members of the Ai2 Climate Modeling team--- Spencer Clark, Brian Henn, Anna Kwa, and Andre Perkins for helpful discussions and software support throughout the development of this work.
Lawrence Livermore National Laboratory is operated by Lawrence Livermore National Security, LLC, for the U.S. Department of Energy, National Nuclear Security Administration under Contract DE-AC52-07NA27344
\bibliography{main}

\end{document}


%
%


\title{Stochastic Emulation of Fully Coupled E3SMv3 Simulation}
%
%

%
%



%
%

%

\begin{article}

%
%

\noindent\textbf{Contents of this file}
\begin{enumerate}
\item Text S1 to S3
\item Tables S1 to S2
\item Figures S1 to S6
\end{enumerate}

\noindent\textbf{Introduction}

In this Supporting Information we include additional details of...

\noindent\textbf{Text S1: Details of stochastic loss and land masking for the ocean component of coupled SamudrACE fine-tuning}

\noindent\textit{Loss definition}

During coupled fine-tuning, both ACE2S and Samudra are optimized using the
same ensemble loss function,
\begin{align}
  \mathcal{L}(F, y)
  = 0.9 \cdot \overline{\mathrm{fCRPS}_{2}(F, y)}
  + 0.1 \cdot \frac{\sqrt{N_m N_k}}{2} \cdot \mathrm{ES}_{\mathrm{SHT}}(F, y),
  \label{eq:ensemble-loss}
\end{align}
where $F$ denotes the model's output distribution for a given output channel
and lead time, realized as an ensemble of $M = 2$ members
$X_1, X_2 \sim F$ generated from the same initial condition; $y$ is the
corresponding target sample from the reference simulation; and
$\overline{\;\cdot\;}$ denotes spatial averaging. 
All fields are normalized per channel to zero mean before the loss is computed. The variance normalization over the training data depends on the component and variable type: ocean and atmosphere diagnostic fields are scaled to unit variance, while atmospheric prognostic variables are scaled by the standard deviation of their temporal tendencies. Finally, the loss is averaged over output channels and summed over lead times.

The first term is the fair continuous ranked probability score
\cite{lang2026aifs}, which for $M = 2$ members evaluates pointwise to
\begin{align}
  \mathrm{fCRPS}_{2}(F, y)
  = \tfrac{1}{2}\left(|X_1 - y| + |X_2 - y|\right)
  - \tfrac{1}{2}\,|X_1 - X_2|.
  \label{eq:fcrps}
\end{align}
This differs from \citeA{perkins2025hiro}, who
used a weighted sum of fair and ``almost fair'' CRPS to guard against the degeneracy that
\citeA{lang2026aifs} attribute to fair CRPS at small ensemble sizes; we
did not observe such degeneracy in this configuration.

The second term is a spectral energy score: a proper scoring rule
\cite{gneiting2007strictly} applied coefficient-wise to the complex
coefficients of the spherical harmonic transform (SHT) of the ensemble
members and the target,
\begin{align}
  \mathrm{ES}_{\mathrm{SHT}}(F, y)
  = \frac{1}{N_\ell N_m} \sum_{\ell, m} w_m \left[
    \tfrac{1}{2}\textstyle\sum_{i=1}^2\big|\hat{X}_{i,\ell m} - \hat{y}_{\ell m}\big|
    - \tfrac{1}{2}\big|\hat{X}_{1,\ell m} - \hat{X}_{2,\ell m}\big|
  \right],
  \label{eq:energy-score}
\end{align}
where $\hat{X}_{i} = \mathrm{SHT}(X_i)$, $|\cdot|$ is the complex modulus,
$N_\ell$ and $N_m$ are the number of retained degrees $\ell$ and orders $m$, and
$w_m = 1$ for $m = 0$ and $w_m = 2$ otherwise (the real SHT stores only
$m \geq 0$; doubling the $m > 0$ modes accounts for their conjugate
counterparts). The $1/(2\sqrt{N_\ell N_m})$ factor brings the term's magnitude
in line with the CRPS term. Because this term scores the distribution of the complex spectral coefficients themselves, evaluating both their real and imaginary components, it directly penalizes ensembles for discrepancies in both spectral power and phase at any wavenumber.

The loss is applied over short stochastic rollouts whose length is sampled
anew for each batch. Fine-tuning windows span four coupled steps (each
coupled step being one 5-day Samudra step, subdivided into twenty 6-hourly
ACE2S steps). For the ocean component the number of optimized forward steps
is drawn from $\{0, 1, 2, 4\}$ with probabilities
$(0.1, 0.3, 0.3, 0.3)$ and the loss is applied at every step of the
sampled rollout; for the atmosphere it is drawn from
$\{0, 1, 2, 4, 21, 41\}$ with probabilities
$(0.025, 0.29, 0.29, 0.29, 0.1, 0.05)$ and only the final step is
optimized. Both components use the loss of
Equation~\eqref{eq:ensemble-loss} with identical weights and unit relative
weighting between the components.

\noindent\textit{Land masking in the loss}

Ocean fields are undefined (NaN) over land; before every loss
term --- including the SHT of the energy score --- land points of both the
prediction and the target are filled with zeros in normalized space, i.e.\
with the global mean of the training data. The resulting sharp
discontinuity at coastlines is expected to inject spurious power at the
highest wavenumbers of both $\hat{X}$ and $\hat{y}$ in
Equation~\eqref{eq:energy-score}, and thus to dilute the energy-score
signal at the smallest scales relative to a smoothly filled field. We
experimented with applying the smooth fill of Text~S2 to the ocean fields
before the energy-score SHT and found similar results, so we have not
pursued it further; a systematic evaluation is left to future work.

\section*{Text S2: Smooth fill of the land surface for ocean spectra}

Ocean fields are undefined over land, but the spherical harmonic transform
used for the power-spectrum diagnostics requires values on the full global
grid. Before computing spectra, land regions are filled by a smooth flood
fill designed to avoid both sharp coastline seams and spurious structure in
the continental interiors. Writing the land region as the set of NaN grid
cells of a given field (constant in time for each variable), the fill has
three phases:

\begin{enumerate}
  \item \textit{Interior mean fill.} Land cells that would not be reached within
        $n$ steps of the edge expansion below (the deep continental interior)
        are identified once per variable by simulating the expansion on the land
        mask with a $3 \times 3$ neighborhood. Each such cell is filled with the
        global mean of the valid (ocean) cells of that map, computed separately
        per sample and time step.
  \item \textit{Edge-blend expansion.} The remaining unfilled land cells (a ring
        of width $n$ along the coastlines) are filled by $n$ iterations of
        neighbor averaging: at each iteration, every unfilled cell adjacent to
        at least one filled or valid cell receives the average of its
        filled/valid $3 \times 3$ neighbors, and is marked filled. This grows
        the ocean values smoothly inland, one cell layer per iteration, until
        they meet the mean-filled interior.
  \item \textit{Gaussian seam smoothing.} A separable Gaussian blur (kernel size
        $k$, standard deviation $\sigma$) is applied to the filled field, and
        the blurred field is blended with the unblurred one using a
        Gaussian-blurred copy of the original ocean mask as the blending weight:
        $x \leftarrow w\,x + (1 - w)\,\tilde{x}$, where $\tilde{x}$ is the
        blurred field and $w$ is the blurred mask (1 in the ocean interior, 0
        deep inland). This leaves ocean values untouched away from the coast
        while smoothly interpolating across the original ocean/land boundary,
        preventing a sharp seam between real and filled data.
\end{enumerate}

All convolutions use circular padding in longitude (periodicity) and
non-periodic padding in latitude (zero for the expansion and mask steps,
edge replication for the Gaussian blur). The diagnostics use the default
parameters $n = 4$ expansion steps, blur kernel size $k = 5$, and
$\sigma = 1$ grid cell. The fill is applied to generated and reference
fields alike, so both spectra in the ratio $r(\ell)$ of Text~S1 are
computed from identically filled fields.

\section*{Text S3: Effect of stochastic training on ocean horizontal spectra}

We quantify the effect of the ensemble loss on the horizontal power spectra
of the ocean fields by comparing the MSE-trained uncoupled Samudra
against the ensemble-loss fine-tuned SamudrACE, each at its best
checkpoint, on two normalized spectral-bias metrics
logged during training, evaluated on the 80 ocean output channels common to
both runs. For each channel, the ratio
$r(\ell) = P_{\mathrm{gen}}(\ell) / P_{\mathrm{ref}}(\ell) - 1$ compares the
generated and reference power at spherical wavenumber $\ell$ (spectra
computed after the smooth fill of Text~S2);
$\overline{|r|}$ (the mean of $|r(\ell)|$ over wavenumbers) measures
whole-spectrum fidelity, and $r(\ell_{\max})$ measures the bias at the
smallest resolved scale. Table~\ref{tab:spectra} reports both metrics for
12-year rollouts started within the training period; the out-of-sample
one-step validation metrics show the same qualitative pattern.

\begin{table}[htbp]
  \centering
  \caption{Normalized spectral bias of ocean channels:\ whole-spectrum
    fidelity $\overline{|r|}$ (smaller is better) and smallest-scale bias
    $r(\ell_{\max})$ (zero is best; negative = too little small-scale
    power), for the MSE-trained uncoupled Samudra and the ensemble-loss
    fine-tuned SamudrACE, each at its best epoch. Values are
    averages over 12-year in-sample rollouts from 8 initial
    conditions; 19-level three-dimensional fields are averaged across
    levels.}
  \label{tab:spectra}
  \begin{tabular}{lrrcrr}
    & \multicolumn{2}{c}{$\overline{|r|}$}
    & & \multicolumn{2}{c}{$r(\ell_{\max})$} \\
    \cmidrule{2-3} \cmidrule{5-6}
    Field & \multicolumn{1}{c}{MSE}
    & \multicolumn{1}{c}{ensemble loss}
    & & \multicolumn{1}{c}{MSE}
    & \multicolumn{1}{c}{ensemble loss} \\
    \midrule
    Sea surface temperature      & 0.084 & \textbf{0.045} & & $-0.103$ & $-0.065$ \\
    Sea surface height           & 0.084 & \textbf{0.057} & & $-0.117$ & $-0.105$ \\
    Potential temperature        & 0.101 & \textbf{0.047} & & $-0.165$ & $-0.061$ \\
    Salinity                     & \textbf{0.036} & 0.038 & & $-0.020$ & $-0.023$ \\
    Zonal velocity               & 0.102 & \textbf{0.072} & & $-0.233$ & $-0.192$ \\
    Meridional velocity          & 0.094 & \textbf{0.090} & & $-0.175$ & $-0.142$ \\
    Sea-ice fraction             & 0.124 & \textbf{0.118} & & $-0.380$ & $-0.329$ \\
    Sea-ice volume per unit area & \textbf{0.031} & \textbf{0.031} & & $-0.066$ & $-0.022$ \\
    \bottomrule
  \end{tabular}
\end{table}

\begin{figure}[htbp]
  \centering\includegraphics[width=0.9\textwidth]{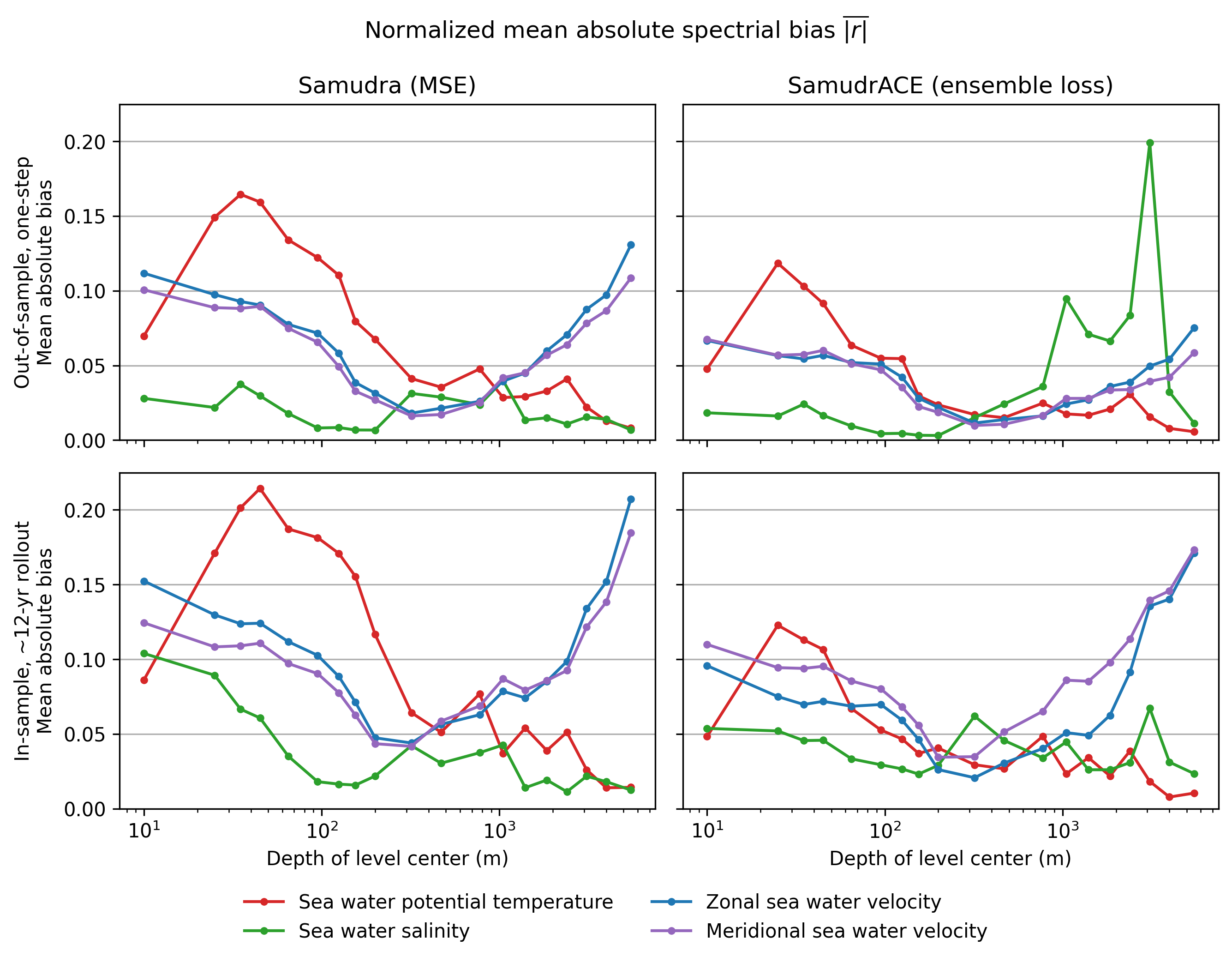}
  \caption{Effect of the stochastic ensemble loss on the horizontal power
    spectra of ocean fields: normalized mean absolute spectral bias metrics of the
    MSE-trained uncoupled Samudra and the ensemble-loss fine-tuned
    SamudrACE at their best checkpoints. With the exception of the one-step spectra of deep ocean salinity, training with the ensemble loss decreases spectral biases throughout the ocean.}
  \label{fig:spectra-depth}
\end{figure}

The MSE-trained Samudra shows too little power at the smallest resolved scales
($r(\ell_{\max})$ is negative for every field, i.e.\ blurring) and is largest in
magnitude for the velocities, potential temperature, and sea ice fraction.
Fine-tuning with the ensemble loss moves the smallest-scale bias toward zero for
almost every field while improving whole-spectrum fidelity for temperature
(level-averaged $\overline{|r|}$ from 0.101 to 0.047), sea surface temperature
(0.084 to 0.045), sea surface height (0.084 to 0.057), and the velocities.
Salinity, with the second smallest magnitude after MSE training, is the one
exception. It is essentially unchanged in the level average (improving in the
upper $\sim$50~m but degrading below $\sim$1000~m, where the fields are very
smooth and the land/floor mask fraction is high). Per-level profiles
(Figure~\ref{fig:spectra-depth}) show the temperature and velocity improvements
are largest in the mesoscale-rich upper $\sim$250~m.

For comparison, within the same coupled run the atmosphere component's spectral
bias is substantially smaller in like-for-like near-interface comparisons (e.g.\
lowest-layer air temperature $\overline{|r|} = 0.016$ versus $0.095$ for 0--40~m
ocean temperature), and the two components sit on opposite sides of zero at the
smallest scales (atmosphere positive, ocean negative). These metrics alone
cannot attribute the asymmetry to the coastline zero-fill in the ocean portion
of the loss, but they hint toward possible room for improvement in the
definition of the loss objective for SamudrACE's ocean component.

\begin{table}
\caption{Variables exchanged between the atmosphere (ACE2S) and ocean (Samudra) emulators during coupled simulation.}
\label{tab:coupling-vars}
\centering
\begin{tabular}{lll}
\hline
Variable & Description & Units \\
\hline
\multicolumn{3}{l}{\textit{Atmosphere $\rightarrow$ Ocean}} \\
\texttt{TAUX}   & Zonal surface wind stress            & N\,m$^{-2}$ \\
\texttt{TAUY}   & Meridional surface wind stress       & N\,m$^{-2}$ \\
\texttt{surface\_precipitation\_rate} & Total surface precipitation rate & kg\,m$^{-2}$\,s$^{-1}$ \\
\texttt{frozen\_precipitation\_rate}  & Frozen precipitation rate & kg\,m$^{-2}$\,s$^{-1}$ \\
\texttt{FLDS}   & Surface downwelling longwave flux    & W\,m$^{-2}$ \\
\texttt{FLUS}   & Surface upwelling longwave flux      & W\,m$^{-2}$ \\
\texttt{FSDS}   & Surface downwelling shortwave flux   & W\,m$^{-2}$ \\
\texttt{FSUS}   & Surface upwelling shortwave flux     & W\,m$^{-2}$ \\
\texttt{LHFLX}  & Surface latent heat flux             & W\,m$^{-2}$ \\
\texttt{SHFLX}  & Surface sensible heat flux           & W\,m$^{-2}$ \\
\hline
\multicolumn{3}{l}{\textit{Ocean $\rightarrow$ Atmosphere}} \\
\texttt{sst}     & Sea surface temperature       & K \\
\texttt{ocean\_sea\_ice\_fraction} & Sea ice fraction & -- \\
\hline
\end{tabular}
\end{table}

\begin{figure}
\centering\includegraphics[width=1\textwidth]{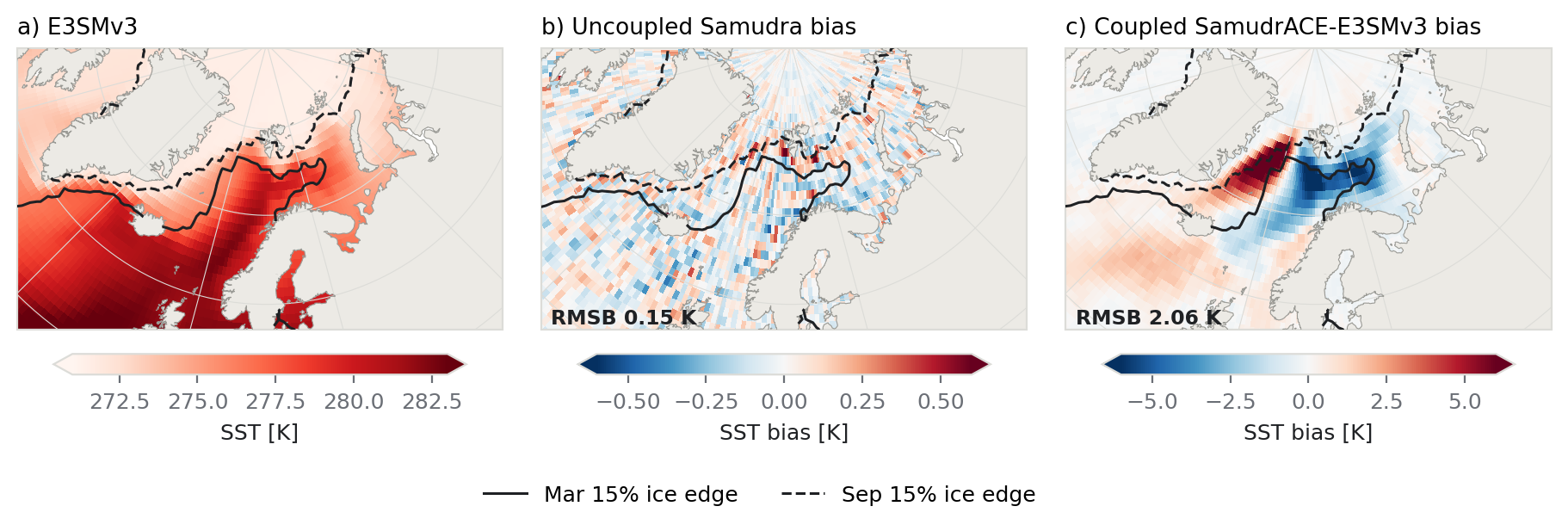}
\caption{Nordic Seas time-mean SST in the 105-yr piControl inference. (a) E3SMv3. (b) Uncoupled Samudra bias. (c) Coupled SamudrACE bias. Contours are the E3SMv3 15\% sea-ice-concentration isolines for March and September in (a) and (c), while (b) uses the uncoupled Samudra's own predicted ice edge. Area-weighted RMSB over the plotted box is annotated on each panel.}
\label{fig:nordic-sst-bias}
\end{figure}

\begin{figure}
\centering\includegraphics[width=1\textwidth]{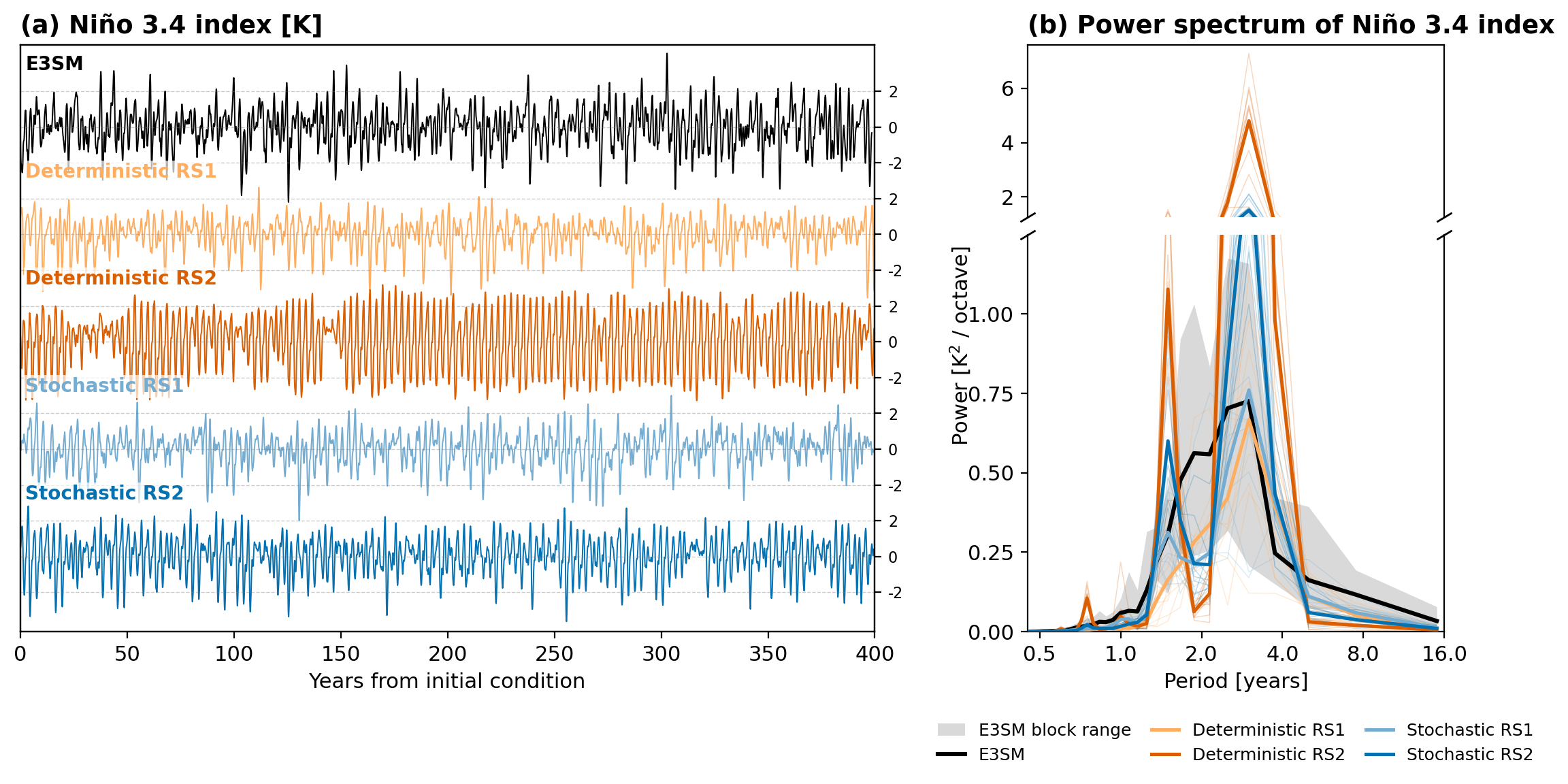}
\caption{Ni\~no~3.4 index (a) and its power spectrum (b) for E3SMv3 and four SamudrACE emulators differing only in training random seed: two deterministic and two stochastic (blue). For each seed we show the checkpoint with the best long-rollout climate skill, selected independently of any ENSO diagnostic. Stochastic RS1 is the model shown in the main text. The index and spectra are computed as in Figure 5; thin lines show individual 40-year blocks, bold lines the block average, and gray shading the range across E3SMv3's blocks. Note the split y-axis in (b): Deterministic RS2's spectral peak is ${\sim}6.6\times$ E3SMv3's. Rollouts here are initialized at the end of the training period and run for 400 years.}
\label{fig:nino34-index-stoc-det}
\end{figure}

\begin{figure}
\centering\includegraphics[width=1\textwidth]{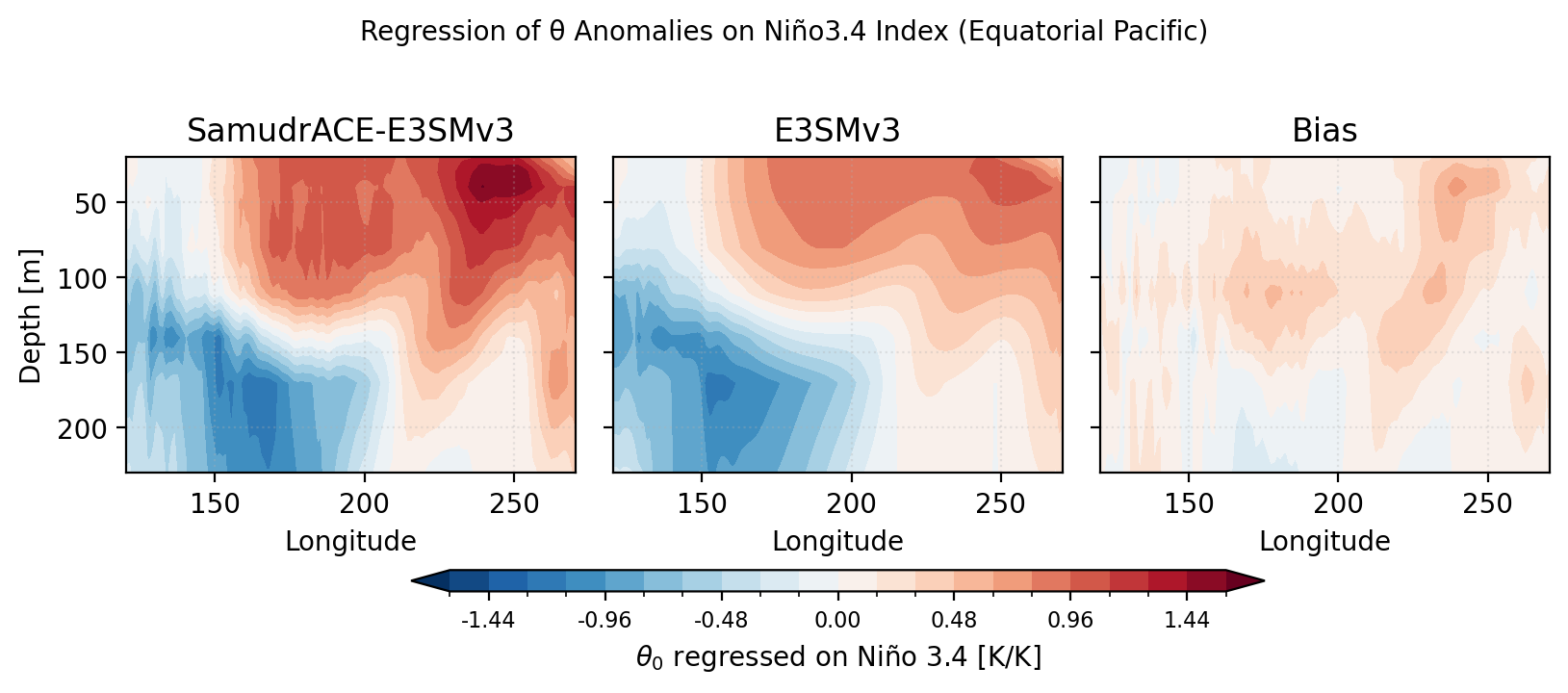}
\caption{Regression of monthly mean ocean temperature ($\theta_o$) anomalies on the Niño 3.4 index for SamudrACE-E3SMv3, E3SMv3, and bias over the 400-year evaluation.}
\label{fig:nino34-thetao-ocean-depth}
\end{figure}

\begin{figure}
\centering\includegraphics[width=0.9\textwidth]{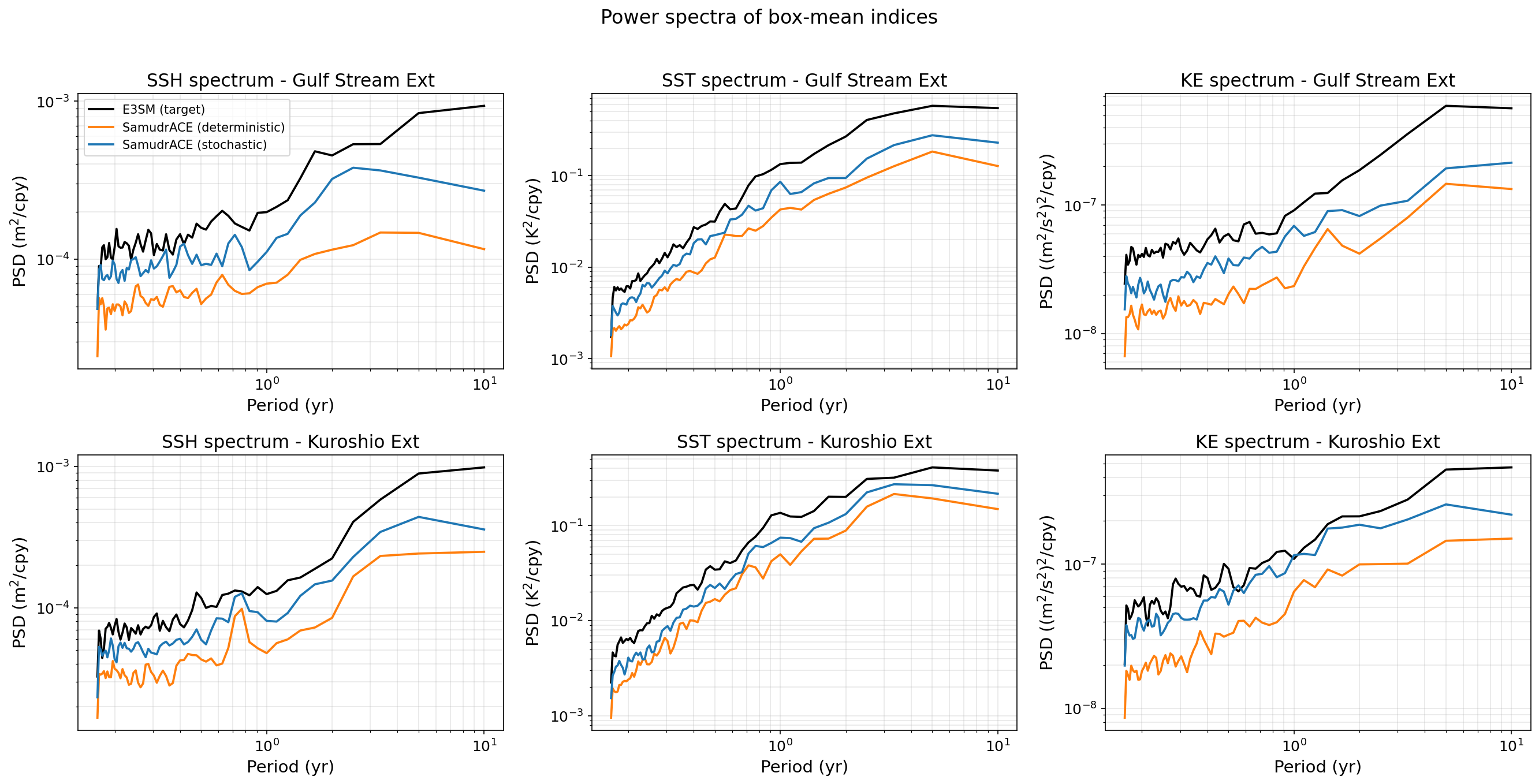}
\caption{Power spectral density of the box-mean anomalies for sea surface height (SSH), sea surface temperature (SST), and surface kinetic energy (KE) in the Gulf Stream and Kuroshio Extension region over the 400-year evaluation for E3SM, deterministic SamudrACE, and stochastic SamudrACE.}
\label{fig:ocn-var-spectra}
\end{figure}

\begin{figure}
\centering\includegraphics[width=0.9\textwidth]{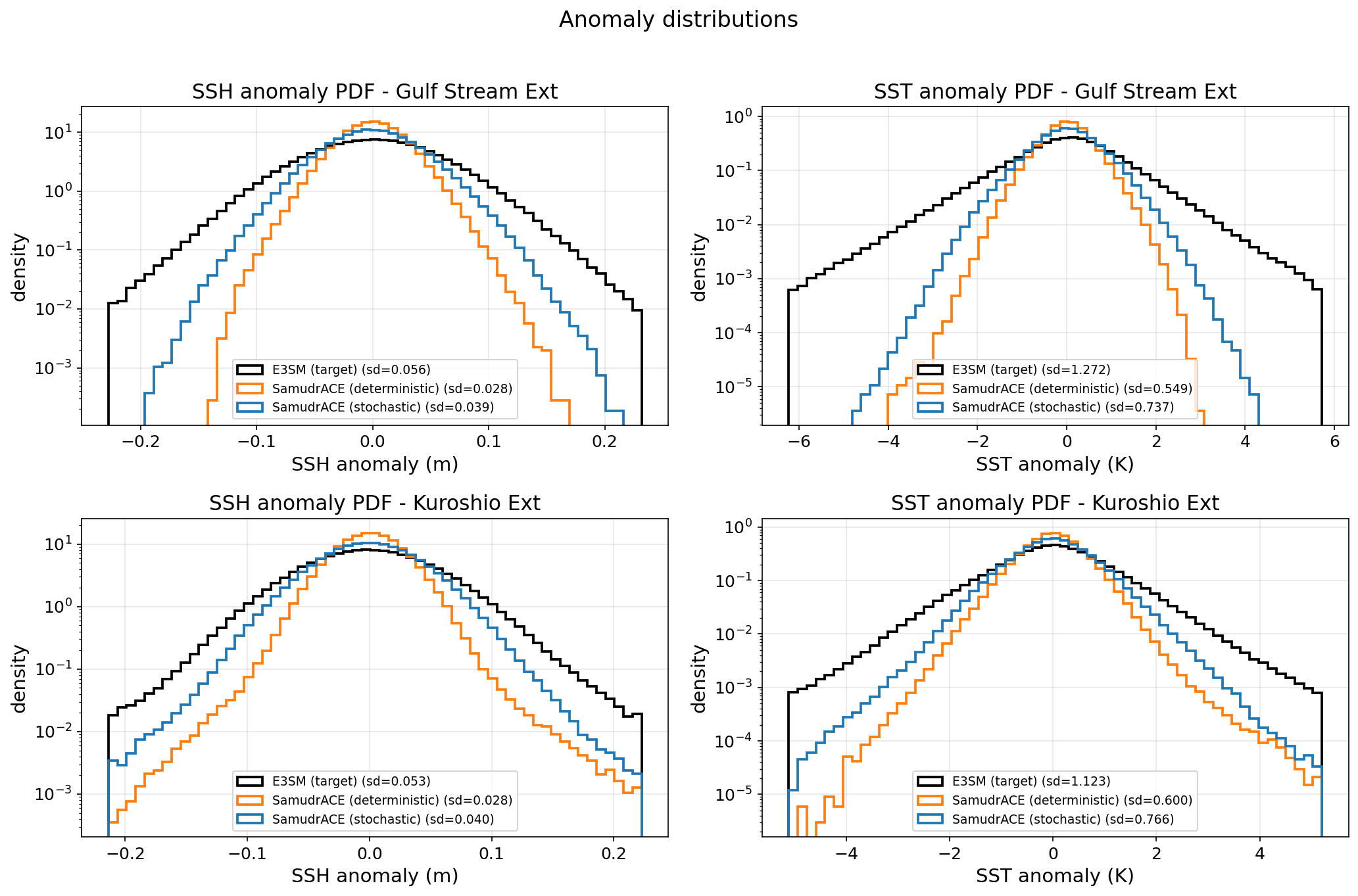}
\caption{Probability density of SSH and SST anomalies in the Gulf Stream and Kuroshio Extension region over the 400-year evaluation for E3SM, deterministic SamudrACE, and stochastic SamudrACE.}
\label{fig:ocn-var-pdf}
\end{figure}

\bibliography{main}

%
%
%
%
%

%
%
\end{article}